\documentclass[aps,pra,superscriptaddress,showpacs,showkeys,floatfix,reprint]{revtex4-1}
\usepackage{graphicx}
\usepackage{dcolumn}
\usepackage[pdftex]{color}
\usepackage{appendix}
\usepackage{amsmath}
\usepackage{caption}
\usepackage{subcaption}
\usepackage{bm}
\usepackage{gensymb}
\usepackage{verbatim}
\definecolor{darkgreen}{rgb}{0.133,0.545,0.133}
\definecolor{orange}{rgb}{1.0,0.76,0.02}

\begin{document}

\title{Unusual composition dependence of transformation temperatures
\newline
in Ti-Ta-X shape memory alloys}

\author{Alberto Ferrari}
\email{alberto.ferrari@rub.de}
\affiliation{Interdisciplinary Centre for Advanced Materials Simulation, Ruhr-Universit{\"a}t Bochum, 44801 Bochum, Germany}
\author{Alexander Paulsen}
\affiliation{Institut f{\"u}r Werkstoffe, Ruhr-Universit{\"a}t Bochum, 44801 Bochum, Germany}
\author{Jan Frenzel}
\affiliation{Institut f{\"u}r Werkstoffe, Ruhr-Universit{\"a}t Bochum, 44801 Bochum, Germany}
\author{Jutta Rogal}
\affiliation{Interdisciplinary Centre for Advanced Materials Simulation, Ruhr-Universit{\"a}t Bochum, 44801 Bochum, Germany}
\author{Gunther Eggeler}
\affiliation{Institut f{\"u}r Werkstoffe, Ruhr-Universit{\"a}t Bochum, 44801 Bochum, Germany}
\author{Ralf Drautz}
\affiliation{Interdisciplinary Centre for Advanced Materials Simulation, Ruhr-Universit{\"a}t Bochum, 44801 Bochum, Germany}

\date{\today}

\begin{abstract}
Ti-Ta-X (X = Al, Sn, Zr) compounds are emerging candidates as high-temperature shape memory alloys (HTSMAs).
The stability of the one-way shape memory effect (1WE), the exploitable pseudoelastic (PE) strain intervals as well as the transformation temperature in these alloys depend strongly on composition, resulting in a trade-off between a stable shape memory effect and a high transformation temperature.
In this work, experimental measurements and first-principles calculations are combined to rationalize the effect of alloying a third component to Ti-Ta based HTSMAs. 
Most notably, an \emph{increase} in the transformation temperature with increasing Al content is detected experimentally in Ti-Ta-Al for low Ta concentrations, in contrast to the generally observed dependence of the transformation temperature on composition in Ti-Ta-X.  
This inversion of trend is confirmed by the \textit{ab-initio} calculations.
Furthermore, a simple analytical model based on the \textit{ab-initio} data is derived.
The model can not only explain the unusual composition dependence of the transformation temperature in Ti-Ta-Al, but also provide a fast and elegant tool for a qualitative evaluation of other ternary systems.
This is exemplified by predicting the trend of the transformation temperature of Ti-Ta-Sn and Ti-Ta-Zr alloys, yielding a remarkable agreement with available experimental data. 
\end{abstract}



\maketitle

\section{Introduction}

\begin{figure}[b]
\begin{centering} 
\includegraphics[scale=0.10]{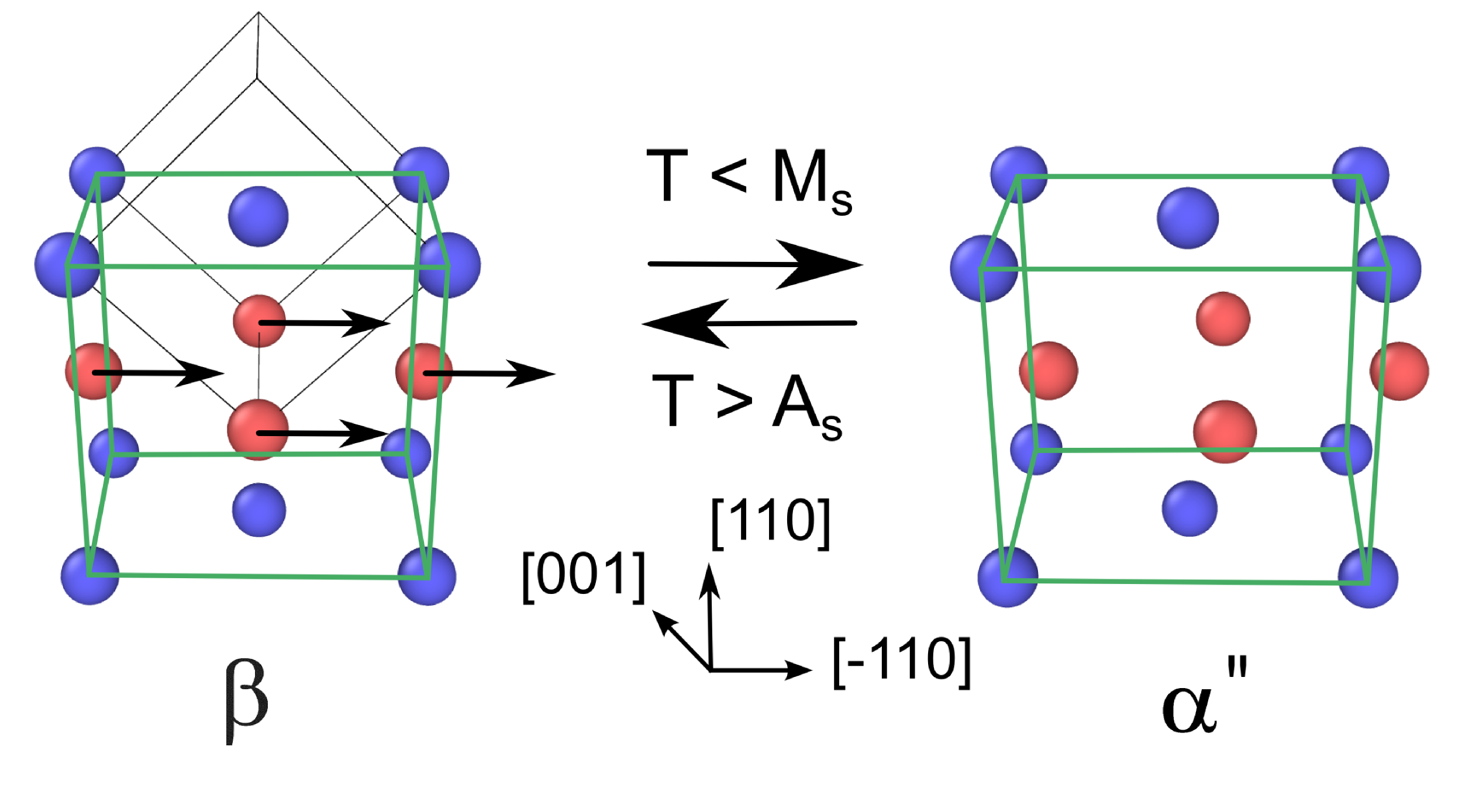}
\par\end{centering}
\caption{The martensitic transformation in $\beta$-Ti based SMAs.  
Black lines indicate the conventional cubic cell of bcc, green lines the conventional orthorhombic cell of the $\alpha''$ phase. 
During the transformation from $\beta$ to $\alpha''$ the lattice vectors are distorted and atoms in red are displaced within the $\left\langle1 1 0\right\rangle$ plane.
}
\label{bet_alp}
\end{figure}
Shape memory alloys~\cite{Chang1951,Buehler1963,Hirose1988,VanHumbeeck2001,Otsuka2002,Ma2010,Jani2014} (SMAs) are functional materials that, after being deformed, are capable of recovering their original shape upon heating or straining.
There are two known shape memory effects (SMEs): the one-way effect (1WE), or thermal memory, and pseudoelasticity (PE),  or mechanical memory. 
The 1WE is exploited in actuator applications, for sensors, coupling devices, fasteners and valves, while PE is exploited in damping applications, highly flexible devices and medical implants like stents~\cite{VanHumbeeck2001,Otsuka2002,Ma2010,Jani2014}.
The two SMEs are based on the diffusionless martensitic transformation between a high-temperature phase (austenite) and a less-symmetric, low-temperature phase (martensite).
Both the 1WE and PE depend on four temperatures that characterize the martensitic transformation; these are the martensite (austenite) start and finish temperatures $M_\text{s}$ and $M_\text{f}$ (respectively $A_\text{s}$ and $A_\text{f}$) that indicate the beginning and end of the formation of the low-temperature (high-temperature) phase upon cooling (heating).
\newline
The commercially most successful shape memory alloy is Ni-Ti~\cite{Buehler1963,Duerig1999,Otsuka1999,Otsuka2005}, owing to its good functional and mechanical properties, its good corrosion resistance and its electric resistance, which allows for direct current heating.
However, binary Ni-Ti shape memory alloys are limited by low $M_\text{s}$ and $A_\text{f}$ (below 100\degree C)~\cite{Buenconsejo2009, Frenzel2010, Frenzel2015}.
The identification of potentially successful high-temperature shape memory alloys (HTSMAs) poses several challenges regarding materials cost, workability, brittleness, phase precipitation and instability of the SME~\cite{Ma2010}.
Furthermore, high-temperature environments enhance the atomic diffusion, which may alter the diffusionless martensitic transformation.
One of the key questions that emerges in HTSMAs design is how the alloy composition influences the martensitic transformation mechanism and the transformation temperatures. 
\newline
$\beta$-Ti based (such as Ti-Ta and Ti-Nb) alloys have been proposed as promising candidates for practical high-temperature applications since they can be easily manufactured into wires or plates~\cite{Bagaryatskii1958,Bywater1972,Fedotov1985,Fedotov1986,Petrzhik1992,VanHumbeeck2001,Buenconsejo2009,Buenconsejo2009a,Buenconsejo2011}.
The 1WE in $\beta$-Ti SMAs is based on the martensitic transformation between the austenitic $\beta$ phase and the martensitic $\alpha''$ phase.
The $\beta$ phase is body-centered cubic (bcc), the $\alpha''$ phase is orthorhombic (space group \textit{Cmcm}) with four atoms positioned ideally at $(0, 0, 0)$, $(0.5, 0.5, 0)$, $(0, 0.6, 0.5)$, and $(0.5, 0.1, 0.5)$,  respectively. 
The martensitic transformation $\alpha'' \rightarrow \beta$ consists of a gliding of a $\left\langle1 1 0\right\rangle$ plane and cell distortion, as depicted in fig.~\ref{bet_alp}.
The shape memory behaviour of $\beta$-Ti alloys is compromised by the formation of the $\omega$ phase~\cite{Frost1954,Hickman1969,Baker1971,Moffat1988}, with a hexagonal structure. 
This detrimental phase forms during quenching (diffusionless formation of the athermal $\omega$ phase) or during ageing at high temperature (diffusive precipitation of the isothermal $\omega$ phase)~\cite{Buenconsejo2009,Niendorf2015}.
Small $\omega$-phase particles represent obstacles to the martensitic transformation $\beta \rightarrow \alpha''$ and consequently lower $M_\text{s}$ and decrease the exploitable transformation strains~\cite{Niendorf2015,Maier2017}.
Moreover, the formation of the $\omega$ phase results in a decrease of the ductility of $\beta$-Ti based alloys, leading to embrittlement and crack initiation~\cite{Buenconsejo2009,Niendorf2015}.
Both experiments~\cite{Buenconsejo2009} and \textit{ab-initio} calculations~\cite{Chakraborty2015} show that the driving force for the formation of the $\omega$ phase decreases with increasing concentration of $\beta$ stabilizing elements.
But the addition of $\beta$-stabilizing elements also results in a decrease of the transformation temperatures.
Alloying Ta to Ti results in a smaller decrease of the transformation temperatures (-20 to -30 K/at.\%)~\cite{Buenconsejo2009,Buenconsejo2011} than the addition of Nb (-40 K/at.\%)~\cite{Kim2006}. Ti-Ta has therefore been identified as the best base alloy for high-temperature applications.
In binary Ti-Ta the formation of the $\omega$ phase is, however, not completely suppressed~\cite{Buenconsejo2009a,Niendorf2015}, unless the Ta concentration is increased to a value where $M_\text{s}$ becomes lower than 100\degree C. 
\newline
A stable 1WE and high transformation temperatures can be achieved by adding a third element to Ti-Ta. 
In recently developed Ti-Ta-X alloys (X = Al, Sn, Zr) the transformation temperatures have been reported to be constant with respect to thermal cycling~\cite{Buenconsejo2009a,Buenconsejo2011,Kim2011,Zheng2013} and higher than 100\degree C.
Furthermore, Ti-Ta-Al and Ti-Ta-Sn alloys have shown a significantly smaller decrease of the transformation temperatures after ageing, which was attributed to a shift of the phase decomposition reaction $\beta \rightarrow \beta + \omega$ to higher temperatures~\cite{Buenconsejo2009a,Niendorf2015}. 
These improved properties make Ti-Ta-Al, Ti-Ta-Sn, and Ti-Ta-Zr the state-of-the-art candidates for $\beta$ Ti-type HTSMAs.
The addition of ternary elements to Ti-Ta does, however, also influence the transformation temperatures~\cite{Buenconsejo2009a,Buenconsejo2011,Kim2011,Zheng2013}.
The substitution of Ta by Al, Sn or Zr results in a decrease of all the transformation temperatures~\cite{Buenconsejo2011,Kim2011,Zheng2013,Yang2015}.
The number of valence electrons and the atomic radius of the alloying element have been identified as the key parameters to characterize the variation of the transformation temperatures~\cite{Buenconsejo2009a}.
\bigskip
\newline
Theoretical work has extensively analysed the binary Ti-Ta system by means of the coherent potential approximation (CPA)~\cite{Soven1967,Gyorffy1972,Li2013} and density functional theory (DFT)~\cite{Dai2012,Chakraborty2015,Chinnappan2016,Barzilai2016,Chakraborty2016}.
The compositional dependence of the transformation temperature has been specifically addressed for Ti-Ta~\cite{Li2013,Chakraborty2015,Chakraborty2016}, Ti-Nb~\cite{Pathak2014,Minami2017} and Ti-Nb-X~\cite{Minami2016,Minami2017,Rajamallu2017}. 
Within CPA  the $\alpha''$ phase is overstabilized~\cite{Li2013} in Ti-Ta for $20\% \lesssim c_\text{Ta} \lesssim 30\%$. 
Using DFT together with the Debye approximation for the vibrational contribution to the free energy, Chakraborty et al.~\cite{Chakraborty2016} have shown that an excellent estimation of $M_\text{s}$ can be achieved from the ground state energy difference and the Debye temperature difference between the $\beta$ and $\alpha''$ phases.
However, in most cases simply the 0~K total energy difference between the two structures $\Delta E^{\beta-\alpha''}$ has been considered~\cite{Li2013,Chakraborty2015,Minami2016,Minami2017} to determine the relative phase stability of the $\alpha''$ and $\beta$ phases, which is strongly correlated with the trend of the transformation temperatures.
Recently, Minami et al.~\cite{Minami2016,Minami2017} have investigated the effect of alloying as many as 46 different elements to Ti-Nb. 
They have found that all elements apart from Sc lower $\Delta E^{\beta-\alpha''}$ at $c_\text{Nb}=12.5\%$, which would indicate a decrease in the transformation temperatures.
Consistently, Rajamallu et al.~\cite{Rajamallu2017} have reported that the addition of Sn and Zr to Ti-Nb alloys stabilizes the $\beta$ phase.
\bigskip
\newline
We combine experimental measurements and electronic structure calculations to achieve a systematic and comprehensive investigation of trends in the transformation temperatures in ternary  Ti-Ta-X alloys with X = Al, Sn, Zr.
One of the key findings of our study is that, in contrast to previous reports, both experiments and simulations reveal an increase in the transformation temperatures for increasing $c_\text{Al}$ in Ti-Ta-Al with low $c_\text{Ta}$.
\newline
To explain the observed change in trend, a simple analytical model is derived based on the \textit{ab-initio} data.
Within the analytical model a qualitative understanding of the composition dependence is achieved in terms of interactions between the different components in the austenite and martensite phases.
The required input data are readily obtained by straightforward \textit{ab-initio} calculations for these rather complex systems, providing a fast and elegant way to screen possible ternary alloys.
In the current study, the effect of alloying Al, Sn and Zr to Ti-Ta on the relative phase stability and the transformation temperatures is thus rationalized using a new model based on the electronic properties of the alloys.
The results of our study serve as a guideline to tailor the Ta and X content aiming for a high temperature shape memory effect in Ti-Ta-X alloys.
\newline
The article is organized as follows: Section~\ref{experimental} presents the experimental setup and the measurements of the transformation temperatures as a function of $c_\text{Ta}$ and $c_\text{Al}$ in Ti-Ta-Al.
Section~\ref{computational} outlines the computational details and examines the interactions between the atomic components, focusing on site preference and ordering effects.
Section~\ref{model} introduces the analytical model for the composition dependence of $\Delta E^{\beta-\alpha''}$ and the transformation temperatures based on the quasi-chemical approach~\cite{Porter2009}.
Section~\ref{dir_fit} discusses the application of this model to the DFT and experimental data on Ti-Ta-Al.
In Section~\ref{binary} the composition dependence of $\Delta E^{\beta-\alpha''}$ and the transformation temperatures are analysed in terms of simple binary interactions which are used to discuss composition effects in Ti-Ta-Al, Ti-Ta-Sn and Ti-Ta-Zr.

\section{Experiments: Material, Methods and Results}\label{experimental}
\begin{figure*}[t]
\begin{centering} 
\includegraphics[scale=0.65]{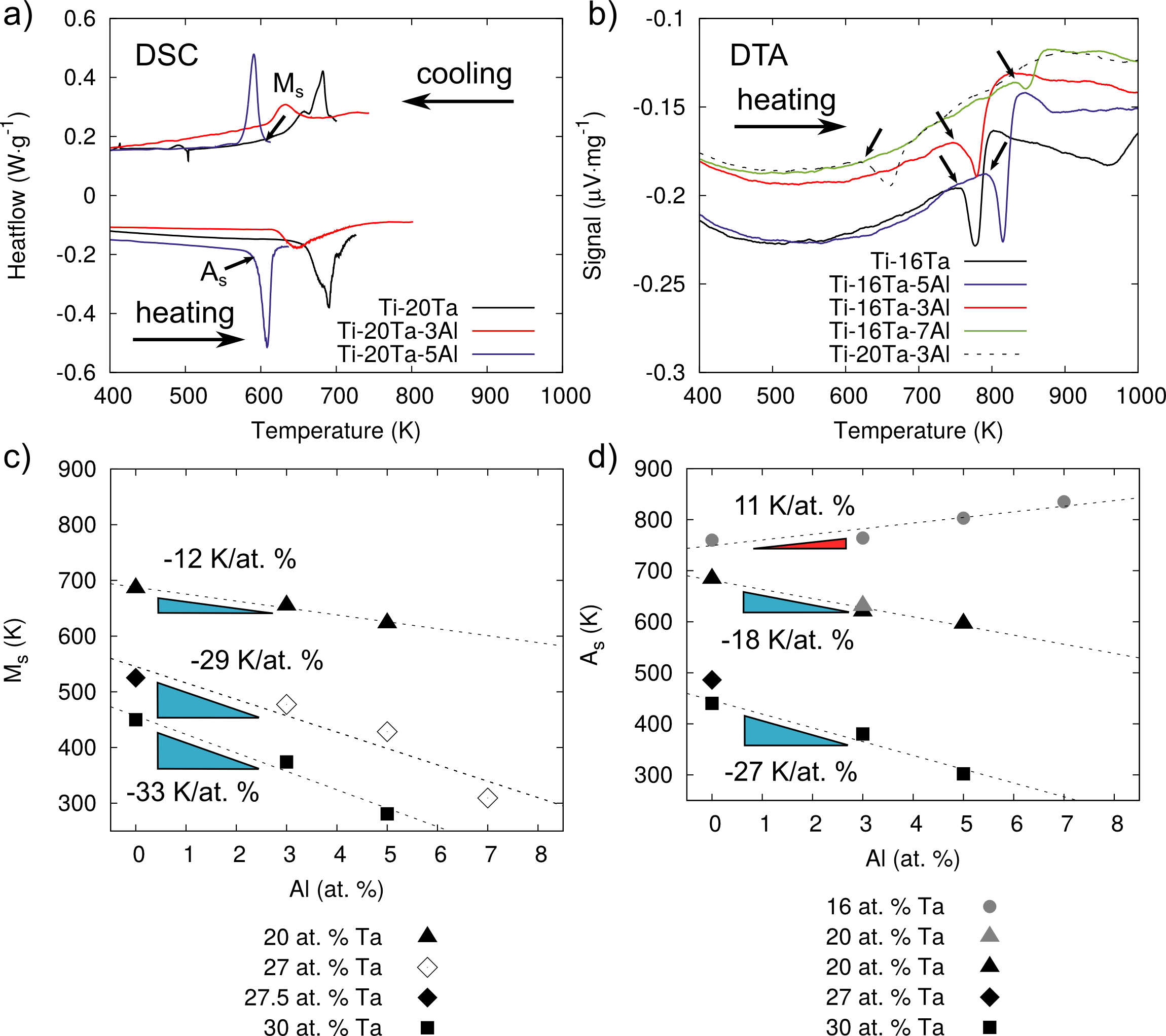}
\par\end{centering}
\caption{Experimental results on phase transformations in Ti-Ta-Al shape memory alloys. \textbf{a)} DSC charts and \textbf{b)} DTA results for different alloy compositions. 
Dependence of \textbf{c)} $M_\text{s}$ and \textbf{d)} $A_\text{s}$ temperatures on alloy composition.
Black symbols represent DSC data, open symbols represent literature data from Buenconsejo et al.~\cite{Buenconsejo2011}, and grey symbols DTA experiments.
\label{experimental_data}
}
\end{figure*}
\subsection{Alloy preparation}
Ti-Ta-Al ingots with masses close to 50 g have been prepared by arc melting under Argon atmosphere using high purity elemental raw materials (Ti: 99.995 wt.\%, Ta: 99.95 wt.\%, Al: 99.99 wt.\%) \cite{materials}.
Details on alloy preparation are documented in~\cite{Zhang2014,Frenzel2010}. 
The as-cast ingots have been homogenized at 1100\degree C for 25 h under high vacuum followed by quenching. 
In a next step, sheets with a thickness of 2 mm were prepared by hot and cold rolling.
The average degree of deformation was close to 0.6. 
The final material has been obtained after cold rolling and subsequent recrystallization annealing at 900\degree C for 10 min followed by water quenching. 
Details on thermomechanical treatments are available in Ref.~\cite{Zhang2014}. 
We have prepared specifically fine-grained sheets since this yields significantly sharper transformation peaks during thermal analysis as compared to coarse-grained ingots.
\subsection{Characterization of phase transformation}
The transformation of the Ti-Ta-Al sheets has been characterized by differential scanning calorimetry (DSC) in a temperature range between room temperature and 600\degree C. 
Differential thermal analysis (DTA) has been used for temperature ranges exceeding 600\degree C \cite{instruments}.
For both thermal analysis techniques, heating rates (and cooling rates in case of DSC) of 20 K/min have been applied.
Further details on thermal analysis are documented in \cite{Allafi2002,Frenzel2015}.
\subsection{Results for Ti-Ta-Al}

Fig.~\ref{experimental_data} summarizes the experimental results on phase transformations as a function of alloy composition in Ti-Ta-Al SMAs. 
Fig.~\ref{experimental_data}a exemplarily shows DSC charts of Ti-20Ta, Ti-20Ta-3Al and Ti-20Ta-5Al.
The peaks on cooling indicate the transformation from $\beta$ to $\alpha''$ martensite. 
Correspondingly, the peaks on heating are related to the reverse transformation. 
Martensite start, $M_\text{s}$, and austenite start, $A_\text{s}$, temperatures are exemplarily highlighted for Ti-20Ta-5Al (blue line in fig.~\ref{experimental_data}b).
A typical DSC experiment started heating from room temperature up to $\approx$50 K above $A_\text{s}$. 
The exposure of the material to high temperatures has been minimized because of the rapid formation of the $\omega$ phase in the austenite, which affects the martensitic transformation upon cooling~\cite{Buenconsejo2009a,Niendorf2015,Maier2017}.
Fig.~\ref{experimental_data}b shows the results of the DTA measurements. 
In contrast to DSC, only the $\alpha'' \to \beta$ transformation during heating is accessible. 
The austenite start temperatures, $A_\text{s}$, are marked with arrows in fig.~\ref{experimental_data}b. 
\newline
The dependence of $M_\text{s}$ and $A_\text{s}$ on the Ta- and Al- concentration as extracted from the DSC and DTA measurements is presented in figs.~\ref{experimental_data}c and d.
The corresponding data on the transformation temperatures are compiled in tab.~\ref{data_transf_temp} in the Appendix.
Black symbols represent DSC data, open symbols represent literature data from Buenconsejo et al.~\cite{Buenconsejo2011}, and grey symbols DTA experiments. 
For Ta concentrations above 20~at.\%, our measurements confirm the general observation that the transformation temperature decreases with both the addition of Ta as well as Al.
However, the effect of alloying Al on $M_\text{s}$ and $A_\text{s}$ clearly decreases for lower Ta concentration. For Ti-30Ta-Al alloys the transformation temperature is reduced by $\sim 27-33$~K/at.\% Al, whereas for Ti-20Ta-Al alloys the addition of Al decreases $M_\text{s}$ and $A_\text{s}$ only by $\sim 12$ and $18$~K/at.\% Al.
For alloys with 16~at.\% Ta the trend is even reversed, and the austenite start temperature \emph{increases} with increasing Al concentration by $\sim 11$~K/at.\% Al.
Due to the high transformation temperatures, DSC experiments could not be performed for the Ti-16Ta-Al alloys.
Since $A_\text{s}$ and $M_\text{s}$ are directly proportional to each other (see Appendix \ref{appendix_exp}), it is expected that $M_\text{s}$ increases at a similar rate for this composition.
\newline
To the best of our knowledge, this is the first report of increasing transformation temperatures with increasing $c_\text{Al}$.
Previous experimental studies on Ti-Ta-Al by Buenconsejo et al.~\cite{Buenconsejo2009a,Buenconsejo2011} have only considered relatively high Ta concentrations and therefore always found that the addition of Al decreases the transformation temperatures.  
For Ti-Nb-Al alloys recent theoretical studies~\cite{Minami2016,Minami2017} have reported a decrease in the energy difference between the $\beta$ and $\alpha''$ phase,  $\Delta E^{\beta-\alpha''}$, which is associated with a decrease in transformation temperature for increasing $c_\text{Al}$ for all investigated Nb concentrations.
To understand how the addition of Al influences the transformation temperature and how this depends on the Ta concentration, ultimately leading to an inversion of the generally observed trend, we have performed a theoretical investigation based on electronic structure calculations.

\section{Modelling Ti-Ta-X alloys}\label{computational}
\subsection{Computational details}\label{comp_det}
The DFT calculations have been performed using the Vienna \textit{Ab-initio} Simulation Package (VASP 5.4)~\cite{Kresse1993,Kresse1996,Kresse1996a} with projector-augmented wave (PAW)~\cite{Bloechl1994,Kresse1999} pseudopotentials including $s$, $p$ and $d$ electrons for the transition metals.
The generalized gradient approximation (GGA) in the original PBE parametrization~\cite{Perdew1996} has been chosen for the exchange-correlation potential. 
The PBEsol~\cite{Perdew2008} functional has also been tested, resulting in no significant discrepancy in total energy differences.
A Monkhorst-Pack grid~\cite{Baldereschi1973,Monkhorst1976} with a Methfessel-Paxton smearing~\cite{Methfessel1989} of 0.05~eV has been employed to integrate the Brillouin zone.
The k-point meshes have been adjusted to the unit cell shape with a linear density of 0.12 2$\pi/$\AA~or less.
Two different  energy cutoffs for the plane wave expansion have been selected in the calculations: 
a high cutoff of 510 eV has been adopted for refined calculations, while a lower cutoff of 410 eV has been used in all other cases.    
The 510 eV cutoff guarantees a numerical convergence of total energy differences less than 1 meV/at., the 410 eV cutoff to within 2 meV/at..
\newline
Ti-Ta forms a solid solution with random occupation of lattice sites by either Ti or Ta.
To simulate the chemical disorder with periodic supercells, special quasirandom structures (SQS)~\cite{Zunger1990,Wei1990} were generated with the Monte Carlo algorithm of a modified version~\cite{Pezold2010,Kossmann2015} of the ATAT package~\cite{VanDeWalle2002}.
The geometrical correlations of pair, 3-body, 4-body, and 5-body figures were included up to  the 9th, 5th, 4th, and 2nd neighbour shells, respectively.
\newline
The equilibrium volumes for the $\alpha''$ and $\beta$ phases were obtained by fitting energy-volume curves with the Birch-Murnaghan equation of state~\cite{Murnaghan1944,Birch1947}.
At each volume, the ionic positions were relaxed until the forces were less than 0.01~eV/\AA.
For the $\alpha''$ phase, the lattice constants ratios $b/a$ and $c/a$ were optimized for every volume.

\subsection{Site preference for alloying elements}\label{results}
\begin{figure}[b]
\begin{centering} 
\includegraphics[scale=0.08]{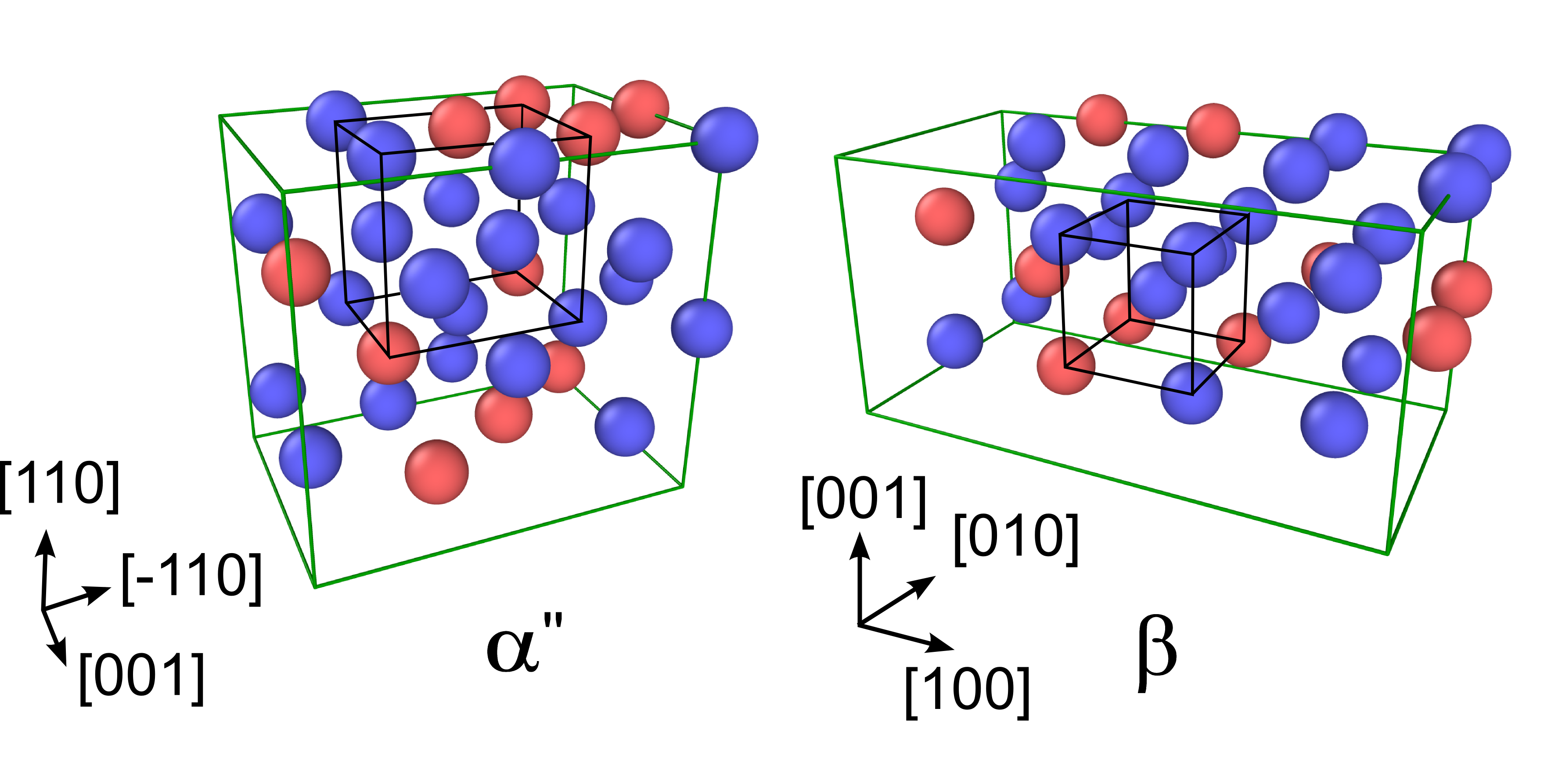}
\par\end{centering}
\caption{SQS structures for $\alpha''$ (left) and $\beta$ (right) at Ti$_{22}$Ta$_{10}$ composition. Ti atoms are blue, Ta atoms are red. Green lines indicate the SQS supercell, black lines indicate the conventional orthorhombic and cubic cells of  the $\alpha''$ and $\beta$ phases, respectively.
\label{SQS_1}
}
\end{figure}

\begin{figure}[t]
\begin{centering} 
\includegraphics[scale=0.50]{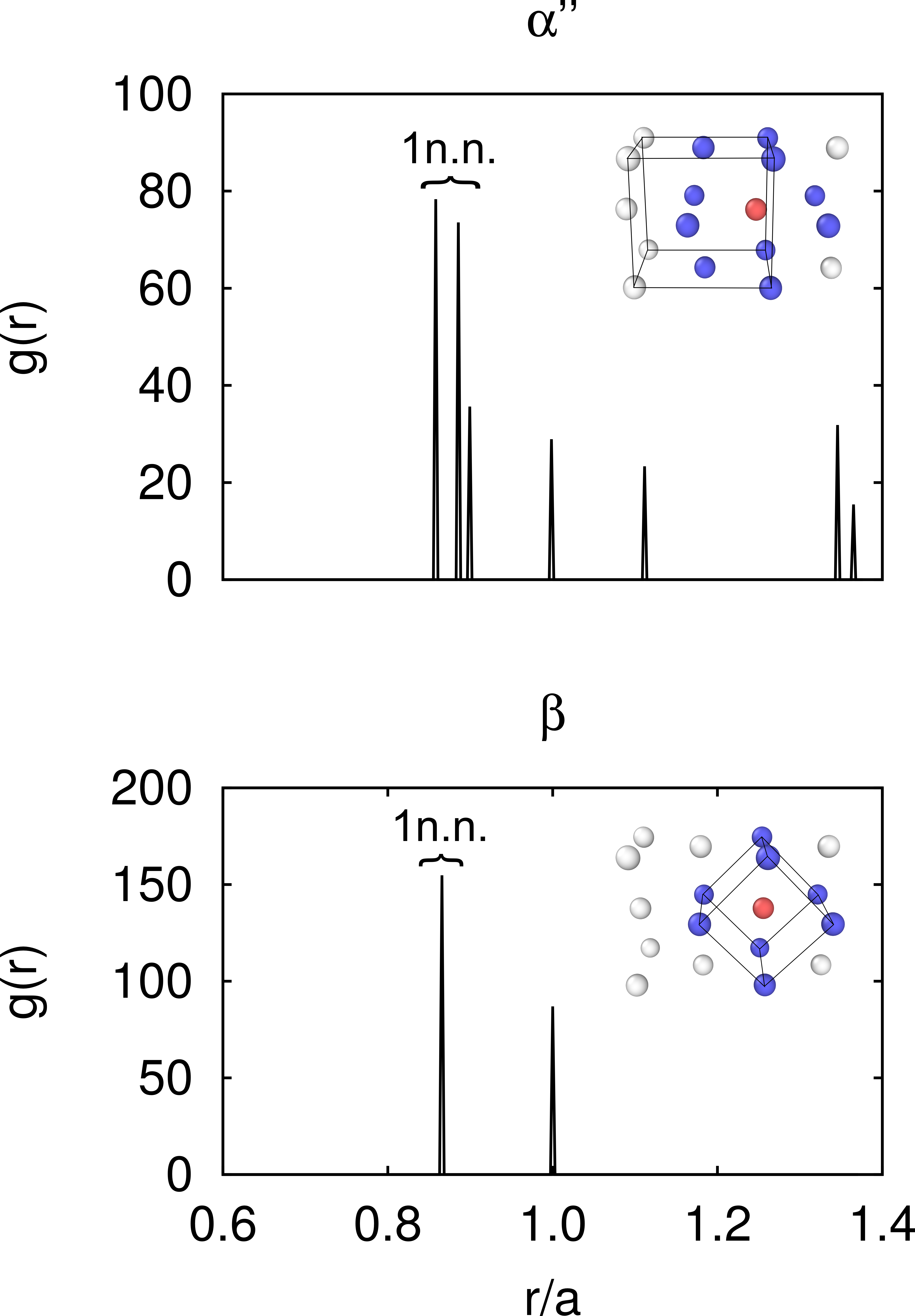}
\par\end{centering}
\caption{Radial distribution function of unrelaxed $\alpha''$ (upper panel) and $\beta$ (lower panel) phases. The distance between atoms is normalized by the lattice parameter $a$ of the two structures. The insets display the nearest neighbours (in blue) of the red atom.}
\label{g_r}
\end{figure}

\begin{figure}[b]
\begin{centering} 
\includegraphics[scale=0.75]{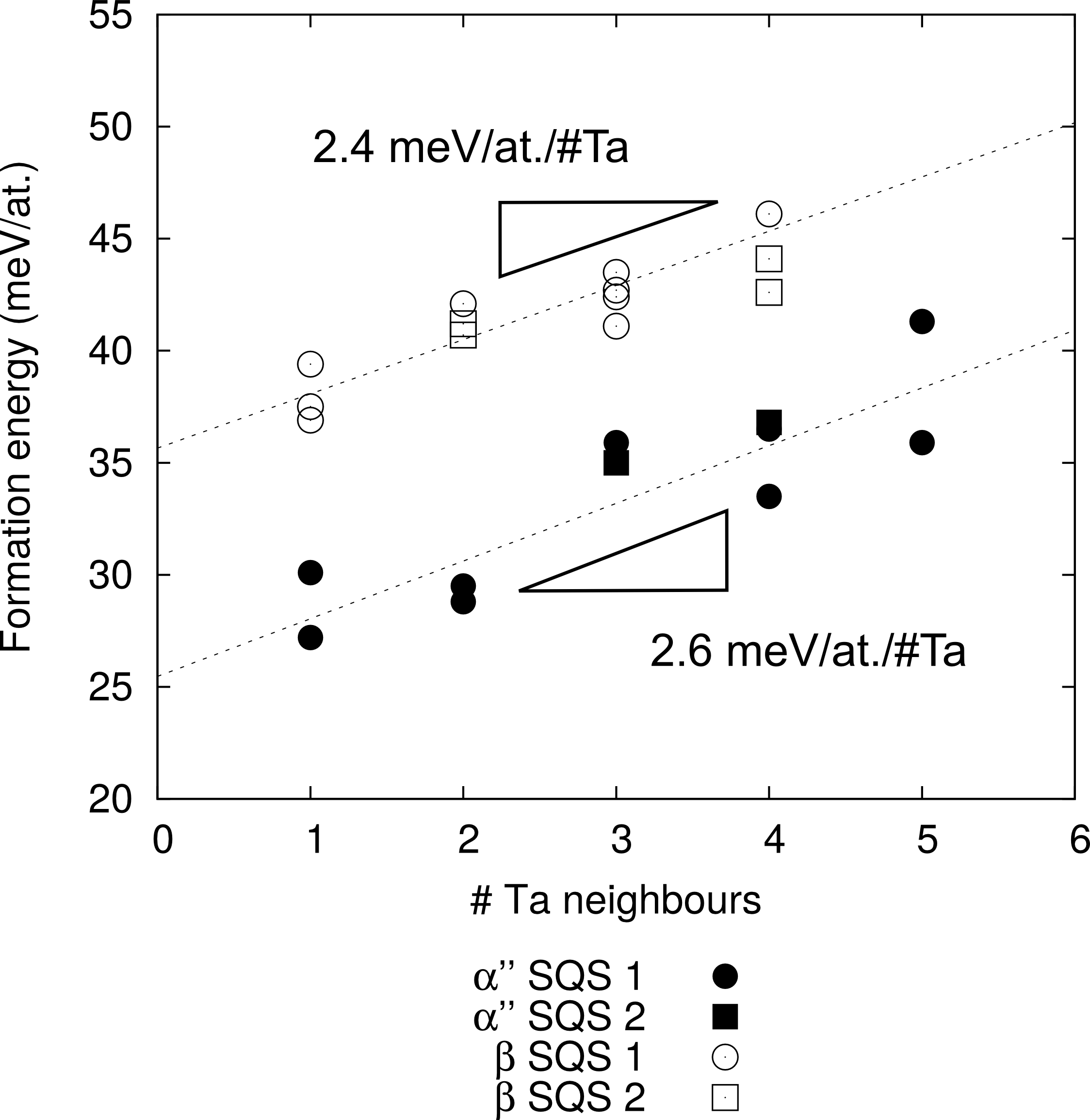}
\par\end{centering}
\caption{Formation energy of Ti$_{22}$Ta$_9$Al as a function of chemical environment. Filled symbols indicate the $\alpha''$ phase, open symbols indicate the $\beta$ phase. Round and square symbols denote different SQS.
\label{Ti-28Ta-3Al}
}
\end{figure}

\begin{figure}[t]
\begin{centering} 
\includegraphics[scale=0.7]{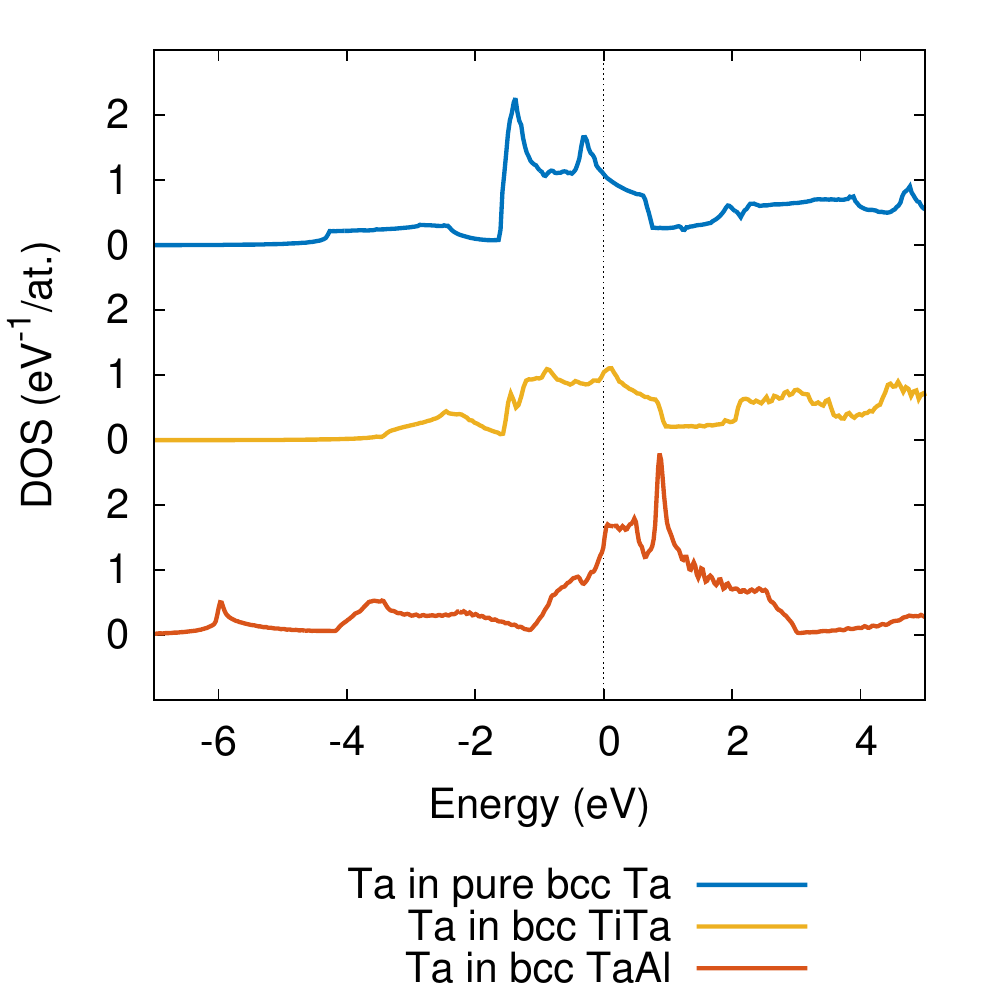}
\par\end{centering}
\caption{Projected density of states for the t$_\text{2g}$ orbitals of Ta in bcc Ta (up), bcc TiTa (middle) and bcc TaAl (low). 
The zero of the energy is the Fermi level.}
\label{dos}
\end{figure}

In order to investigate the effect of adding Al to Ti-Ta, 32-atom SQS of Ti-Ta have been prepared with 31.25\% Ta (Ti$_{22}$Ta$_{10}$) for $\alpha''$ and $\beta$ (see fig.~\ref{SQS_1}).
The optimized lattice parameters for the $\beta$ ($a = 3.28$ \AA) and $\alpha''$ ($a= 3.33$ \AA, $b=4.76$ \AA, $c=4.46$ \AA) phases are in good agreement with previously reported data~\cite{Chakraborty2016}.
One Ta atom in the SQS has subsequently been substituted by an Al atom, resulting in a Ti$_{22}$Ta$_9$Al alloy.
After relaxation of forces and lattice parameters the formation energy of the alloy has been calculated with the high-accuracy setting
\begin{equation}
E^{\text{(i)}}_f = E^{\text{(i)}} - c_\text{Ti} E_\text{Ti} - c_\text{Ta} E_\text{Ta}- c_\text{Al} E_\text{Al} \quad ,
\label{formation_energy}
\end{equation}
where $E^{\text{(i)}}=E^{\text{(i)}}(c_\text{Ti}, c_\text{Ta}, c_\text{Al})$ is the total energy of Ti-Ta-Al in the $i$-th phase ($\beta$ or  $\alpha''$) and here $E_\text{Ti}$ is the total energy of hcp Ti, $E_\text{Ta}$ of bcc Ta and $E_\text{Al}$ of fcc Al.
\newline
Substitutions have been performed at Ta sites with different chemical environment (i.e.~different number of Ti and Ta neighbours).
The corresponding formation energies are correlated with the number of Ta atoms in the nearest neighbour (n.n.)~shell surrounding the Al impurity.
All atoms up to the first gap in the radial distribution function (RDF) are considered to be in the first n.n.~shell.
For the $\alpha''$ phase this includes the first three peaks of the RDF shown in fig.~\ref{g_r} with a total number of 10 atoms, for the $\beta$ phase this corresponds to the 8 first n.n.~in the first peak of the RDF.
The formation energies of the different configurations as a function of the number of Ta atoms in the n.n.~shell is shown in fig.~\ref{Ti-28Ta-3Al}. 
The $\alpha''$ phase is the ground state phase for Ti$_{22}$Ta$_9$Al, in agreement with experimental findings~\cite{Buenconsejo2009a,Buenconsejo2011} at similar compositions.
For different configurations with the same number of Ta n.n.~the formation energies vary up to $\sim 5$~meV/at. due to different occupations of more distant neighbour sites.
The overall trend in the formation energies of both phases shows an increase as the number of Ta atoms around the Al impurity increases, which indicates a driving force towards ordering.
The driving force originates from the fact that Ta atoms can only accommodate the $p$-valent Al defects by reorganizing their t$_\text{2g}$ orbitals, which is energetically unfavourable in $\beta$ and $\alpha''$ symmetries.
This can be deduced by comparing the projected densities of states of the t$_\text{2g}$ orbitals of Ta in bcc Ta, bcc TiTa, and bcc TaAl shown in fig.~\ref{dos}.
The electronic structure of Ta undergoes substantial changes around the Fermi level when Ta binds to Al with respect to the more favourable situation, when Ta binds to Ta or Ti. 
As a result, the effective interaction between Ta and Al is repulsive.
Due to the repulsive interaction between Ta and Al, Ti-Ta-Al would not be rigorously a random alloy at $T=0$~K, since Al will try to minimize the number of Ta atoms in the n.n.~shell.
The martensitic transformation is, however, a displacive transformation and it is unlikely that atoms are able to diffuse to find the lowest energy configuration at low temperatures~\cite{Hennig2005}. 
Furthermore, the driving force towards ordering is small for low defect concentrations.
In fact, in the range of compositions relevant to SMAs, the formation energy difference between $\beta$ and $\alpha''$ calculated using a random occupation is only 3-5 meV/at.~higher than the formation energy difference calculated using the ground state occupation.
This means that, although being detectable by the present DFT calculations, the effect of ordering around Al atoms is negligible in terms of formation energy differences.
\newline
A similar set of DFT calculations has been performed for Ti-Ta-Sn and Ti-Ta-Zr at the same composition.
Ti-Ta-Sn exhibits a very similar behaviour of the formation energy of the two phases as a function of the number of Ta n.n.. 
For Ti-Ta-Zr only the formation energy of the $\beta$ phase is found to increase for increasing number of Ta n.n., while the formation energy of the $\alpha''$ phase in this compound does not depend on the number of Ta in the n.n.~shell.
This may be due to the fact that Zr is isoelectronic to Ti and therefore it binds easily to Ta in the $\alpha''$ phase. 
Consequently, the effect of ordering can as well be neglected in Ti-Ta-Sn and Ti-Ta-Zr.
\newline
Overall the calculations show that Ti-Ta-X alloys can be well approximated as random alloys (regular solid solutions) for low X concentrations $c_{X}$.
\newline
To test the impact of the choice of a particular SQS for the calculations, a second set of SQS (SQS 2 in fig.~\ref{Ti-28Ta-3Al}) was used  for $\alpha''$ and $\beta$ structures using Ti$_{23}$Ta$_9$ as initial composition.
In this case, one Ti atom in the SQS has been substituted by Al, resulting in the same Ti$_{22}$Ta$_9$Al composition.
The formation energies are shown in fig.~\ref{Ti-28Ta-3Al}. 
Both SQS exhibit the same trend in the formation energy difference between $\beta$ and $\alpha''$ as well as in the dependence on the number of Ta neighbours around the Al impurity.
The systematic error introduced by the choice of a specific SQS is less than 5 meV/at.

\section{Composition dependence of the transformation temperature} \label{model}

For Ti-Ta the trend of the four transformation temperatures as a function of composition can be captured quantitatively by the trend in the phase transition temperature $T_0$ between the $\beta$ and $\alpha''$ phase~\cite{Chakraborty2016}, which is the temperature at which the free energies of the two phases are equal.
Instead of evaluating the full free energy as a function of $c_\text{Ti}$, $c_\text{Ta}$ and $c_\text{X}$, which requires a significant computational effort, the relative phase stability is often sufficiently approximated by the difference in formation energies between the two phases
\begin{equation}
\Delta E^{\beta-\alpha''} (c_\text{Ti}, c_\text{Ta}, c_\text{X}) = E_f^{(\beta)} - E_f^{(\alpha'')}
\label{energy_diff}
\end{equation}
Consistently with previous studies~\cite{Chakraborty2015,Frenzel2015,Li2013,Minami2016,Minami2017}, it is assumed that the compositional dependence of $\Delta E^{\beta-\alpha''}$ follows the one of the transition temperature $T_0$.
This approximation is equivalent to stating  that the difference in the entropic terms for the two phases does not depend on $c_\text{Ti}$, $c_\text{Ta}$ and $c_\text{X}$.
\newline
The formation energies for the $i$-th phase ($i=\alpha''$, $\beta$) in eq.~\eqref{energy_diff} are calculated according to eq.~\eqref{formation_energy}, which is the standard definition of formation energies.
In addition, a mixing energy can be defined as
\begin{equation}
^\text{mix}E^{\text{(i)}} = E^{\text{(i)}} - c_\text{Ti} E_\text{Ti}^{\text{(i)}} - c_\text{Ta} E_\text{Ta}^{\text{(i)}} - c_\text{X} E_\text{X}^{\text{(i)}}
\label{mixing}
\end{equation}
where $E_\text{Ti}^{\text{(i)}}$, $E_\text{Ta}^{\text{(i)}}$ and $E_\text{X}^{\text{(i)}}$ indicate the total energies of the pure elements in the $i$-th phase ($i=\alpha''$, $\beta$).
$^\text{mix}E^{(i)}$ from eq.~\eqref{mixing} and $E^{(i)}_f$ from eq.~\eqref{formation_energy} are related to each other through the energy difference between the $i$-th phase and the ground state phase of the pure elements.
For a random alloy, the lowest order approximation for the mixing energy is of second order in the concentrations
\begin{equation}
 ^\text{mix}E^\text{(i)}\simeq k^\text{(i)}_\text{TiTa} c_\text{Ti} c_\text{Ta}+ k^\text{(i)}_\text{TiX} c_\text{Ti} c_\text{X} + k^\text{(i)}_\text{TaX} c_\text{Ta} c_\text{X}
\label{random_all}
\end{equation}
This approximation may be derived for a pairwise cluster expansion \cite{Sanchez1984} in a random alloy and is also referred to as quasi-chemical treatment of regular solutions \cite{Porter2009}.
Using eq.~\eqref{random_all}, eq.~\eqref{energy_diff} can be approximated to second order as
\begin{equation}
 \Delta E^{\beta-\alpha''} \simeq \sum_\text{n}\lambda_\text{n}c_\text{n} + \sum_{\text{n}\neq\text{m}}\Delta k_\text{nm}c_\text{n}c_\text{m}
 \label{II_order}
\end{equation}
where $\lambda_\text{n}=E_\text{n}^{(\beta)}-E_\text{n}^{(\alpha'')}$, $\Delta k_\text{nm}:=k^{(\beta)}_\text{nm}-k^{(\alpha'')}_\text{nm}$ and $\{\text{n, m}\}=\{\text{Ti, Ta, Al}\}$.
The first term in eq.~\eqref{random_all} can be neglected ($k^\text{(i)}_\text{TiTa} \simeq 0$), because Ti-Ta has a very small mixing energy for both phases (i.e.~Ti-Ta alloys can be well approximated by ideal solid solutions).
Indeed, Ti and Ta differ by only one valence electron and their atomic radii are similar, hence the heat of formation of random Ti-Ta compounds is small. 
This is also shown numerically in Section \ref{binary}.
Setting $c_\text{Ti} = 1 - c_\text{Ta}- c_\text{X}$, the number of unknown coefficients in eq.~\eqref{II_order} can be further reduced
\begin{equation}
\Delta E^{\beta-\alpha''} \simeq \text{A} \cdot c_\text{Ta} + \text{B} \cdot c_\text{X}+ \text{C} \cdot c_\text{Ta} c_\text{X} + \text{D}
\label{fitting_function}
\end{equation}
where terms proportional to $c_\text{X}^2$ have also been neglected, since the focus is on small X concentrations ($c_\text{X}\lesssim10\%$).
\newline
The linear dependence of $\Delta E^{\beta-\alpha''}$ on  $c_\text{Ta}$ in eq.~\eqref{fitting_function} has  already been pointed out in previous studies~\cite{Buenconsejo2009, Buenconsejo2011, Kim2011, Chakraborty2015, Chakraborty2016} and it is due to the low heat of formation of Ti-Ta.
Eq.~\eqref{fitting_function} is the simplest formulation of the composition dependence of $\Delta E^{\beta-\alpha''}$ for Ti-Ta-X alloys and respectively of the transition temperature $T_0$, as discussed above.
It can be used to fit DFT or experimental data and it provides an interpretation of the values for the fitting parameters in terms of interatomic interactions.
\newline
Despite having been derived for Ti-Ta-X alloys, the same equation can be used to describe the composition dependence of the transformation temperatures in other random $\beta$-Ti ternary SMAs, provided that the underlying approximations are verified.
Furthermore, this approach  could be trivially extended to quaternary and other multicomponent alloys.

\section{DFT results for Ti-Ta-Al}\label{dir_fit}
\begin{figure}[h]
\begin{centering} 
\includegraphics[scale=0.75]{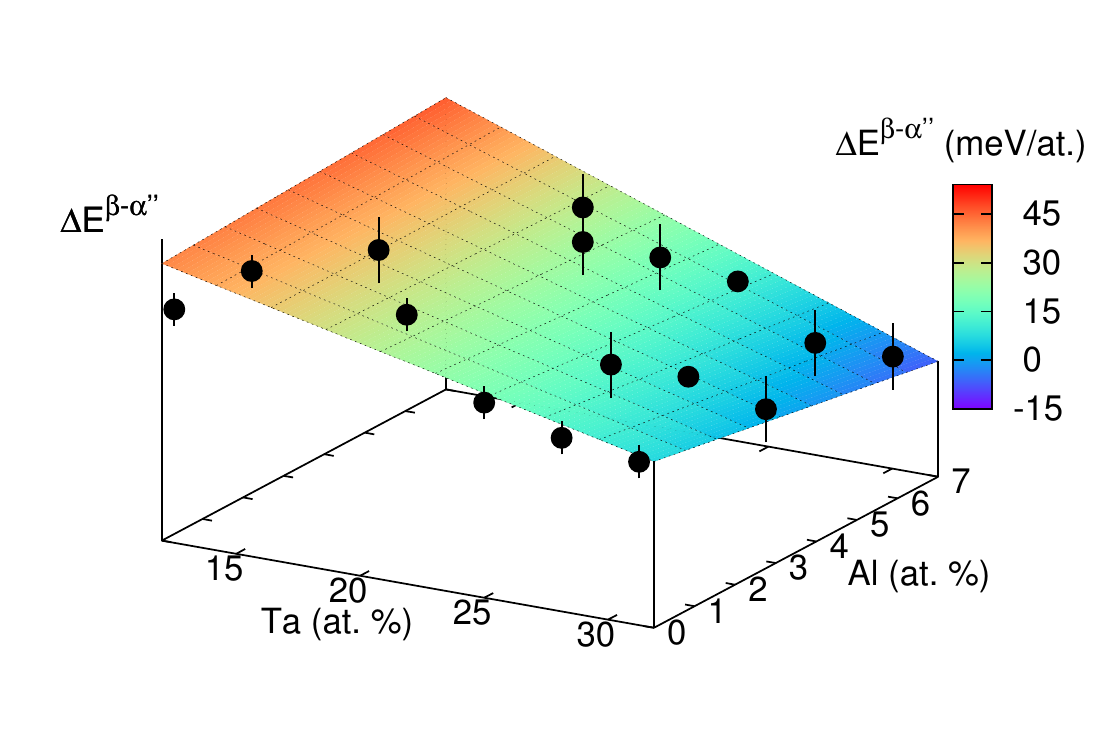}
\par\end{centering}
\caption{Difference of formation energies between the $\alpha''$ and $\beta$ phases as a function of composition, along with best fit of the data.}
\label{DFT_TiTaAl}
\end{figure}

\begin{figure}[h!]
\begin{centering} 
\includegraphics[scale=0.75]{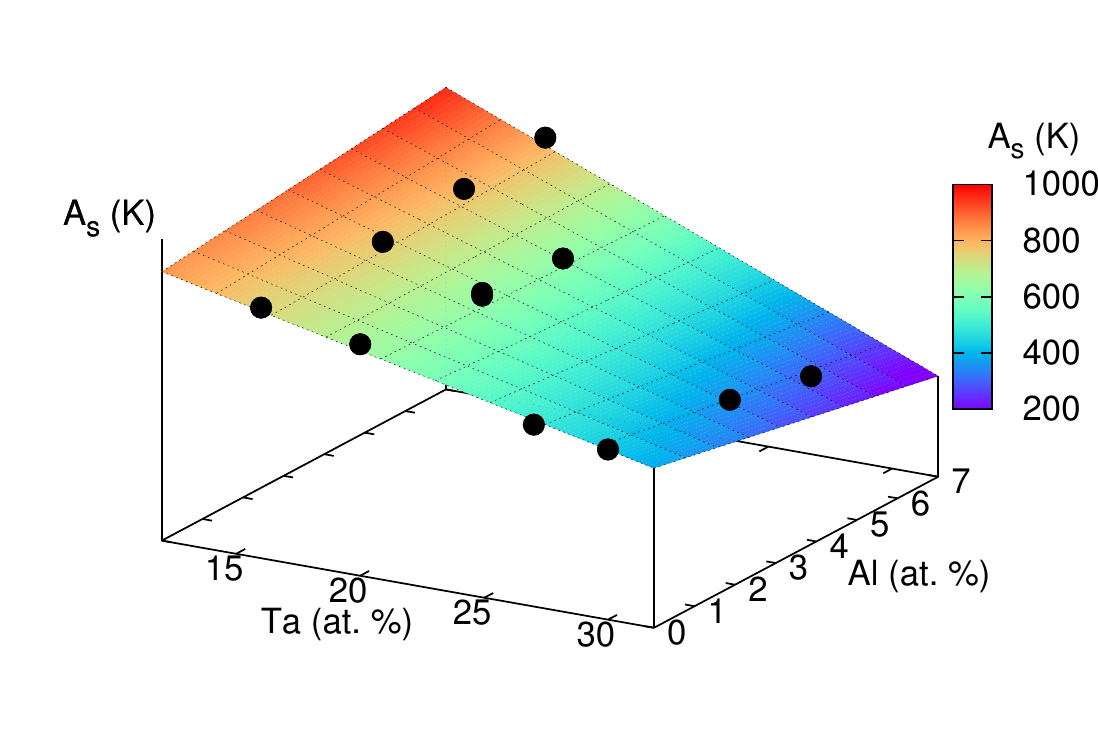}
\par\end{centering}
\caption{Austenite start temperature as a function of composition. The color gradient has been adjusted to be the same as that of fig.~\ref{DFT_TiTaAl}.}
\label{exp_TiTaAl}
\end{figure}
For Ti-Ta-Al $\Delta E^{\beta-\alpha''}$ has been calculated with DFT for 16 different Ta and Al compositions.
The configurations have been setup in two different ways:
\begin{itemize}
\item{for Ti$_{22}$Ta$_9$Al and Ti$_{22}$Ta$_8$Al$_2$, Al has been substituted in several positions of the corresponding SQS (as in Section \ref{results}) and $\Delta E^{\beta-\alpha''}$ has been calculated with a random occupation of nearest neighbours sites}
\item{for all other concentrations, Al has been substituted randomly in the $\beta$ phase and the corresponding $\alpha''$ phase has been considered according to the orientation relationship shown in fig.~\ref{bet_alp}}
\end{itemize}
The high-accuracy setting has been employed for Ti$_{22}$Ta$_9$Al, Ti$_{22}$Ta$_8$Al$_2$, Ti$_{22}$Ta$_{10}$ and Ti$_{28}$Ta$_4$, whereas the lower energy cutoff has been employed for the other compositions. 
Pure Ti-Ta has been assigned an error of 5 meV/at., arising mainly from the SQS approximation, while for Ti-Ta-Al statistical errors of 1 meV/at.~and 10 meV/at.~have been assumed  for the first  and second kind of data, respectively, due to the effect of the number of Ta n.n.~discussed in the previous Section.
Fig.~\ref{DFT_TiTaAl} presents $\Delta E^{\beta-\alpha''}$ as a function of $c_\text{Ta}$ and $c_\text{Al}$ including the corresponding error bars.
The corresponding values are compiled in tab.~\ref{data_DFT} in the Appendix.
The data have been fitted using eq.~\eqref{fitting_function} with a least-squares algorithm.
The fit is acceptable by means of the $\chi^2$ test with a reduced $\chi^2$ of 2.87.
The fitted coefficients are reported in the first column of tab.~\ref{fit_coeff} with their statistical error.
The fitted A coefficient is negative, $-19.6\pm0.8$~K/at.\%, meaning that the addition of Ta stabilizes the $\beta$ structure (due to $d$-band filling) with respect to the $\alpha''$ structure.
Its value is not affected by the addition of Al and it is in good agreement with the value of -24.4 K/at.\% from a previous study on Ti-Ta~\cite{Chakraborty2016}.
The coefficient B in tab.~\ref{fit_coeff} is positive, indicating that when Al substitutes Ti in the compound the $\alpha''$ phase is stabilized more than the $\beta$ phase.
Conversely, the C coefficient is negative, which means that the Ti-Al and Ta-Al interactions tend to favour the $\beta$ phase over $\alpha''$.

\setlength{\tabcolsep}{2pt}
\renewcommand{\arraystretch}{1.5}
\begin{table}[t]
\begin{centering} 
\begin{tabular}{cccc}
\hline
 & DFT & Expt. ($M_\text{s}$) & Expt. ($A_\text{s}$) \tabularnewline
\hline
A (K/at.\%) & -19.6$\pm$0.8 & -22.6$\pm$3.1 & -21.9$\pm$2.5 \tabularnewline
B (K/at.\%) & 27$\pm$16 & 29$\pm$24 & 50$\pm$15 \tabularnewline
C (K/(at.\%)$^{2}$) & -1.7$\pm$0.6 & -2.1$\pm$0.9 & -2.6$\pm$0.7 \tabularnewline
D (K) & 720$\pm$20 & 1140$\pm$80 & 1100$\pm$60 \tabularnewline
\hline
\end{tabular}
\par
\end{centering}
\caption{Fitted coefficients from eq.~\eqref{fitting_function} for $\Delta E^{\beta-\alpha''}$ (DFT) and $A_\text{s}$ and $M_\text{s}$ (Expt.) with statistical errors.
The conversion factor used to compare energy differences and temperatures is $1/k_\text{B}$, where $k_\text{B}$ is the Boltzmann constant.
\label{fit_coeff} 
}
\end{table}

The DFT data in fig.~\ref{DFT_TiTaAl} show qualitatively the same trend in $\Delta E^{\beta-\alpha''}$ as the experimental data for $A_\text{s}$ shown in fig.~\ref{exp_TiTaAl}.
For relatively high ($\simeq$ 30\%) Ta content $\Delta E^{\beta-\alpha''}$ decreases with increasing $c_\text{Al}$, whereas for low ($\lesssim$ 15\%) Ta concentrations $\Delta E^{\beta-\alpha''}$ \emph{increases} with increasing $c_\text{Al}$. 
From eq.~\eqref{fitting_function} it follows that this inversion in trend is due to $\text{B}>0$ and $\text{C}<0$.
The fit predicts the inversion of  trend in $\Delta E^{\beta-\alpha''}$ at $c_\text{Ta}\simeq$ 16\%, which is in very good agreement with the inversion point determined experimentally (at 16\% $<c_\text{Ta}<$ 20\% for $A_\text{s}$).
The corresponding coefficients obtained by fitting  the experimental data on $M_\text{s}$ and $A_\text{s}$ using eq.~\eqref{fitting_function} are also reported in tab.~\ref{fit_coeff}.  
The overall agreement between the theoretical and experimental results is remarkable.
The coefficient D is affected by a constant shift because of the lack of vibrational contributions in the DFT data and as a consequence the value of this coefficient is not fully comparable to the experimental data.
For all other coefficients  the fitted values from the DFT and experimental data agree within one standard deviation or less.
Therefore, the exceptional inversion of the trend in the transformation temperatures as a function of composition can indeed be explained by considering the formation energy difference between austenite and martensite.

\section{Binary interactions method}\label{binary}

\begin{figure*}
\begin{centering} 
\includegraphics[scale=0.5]{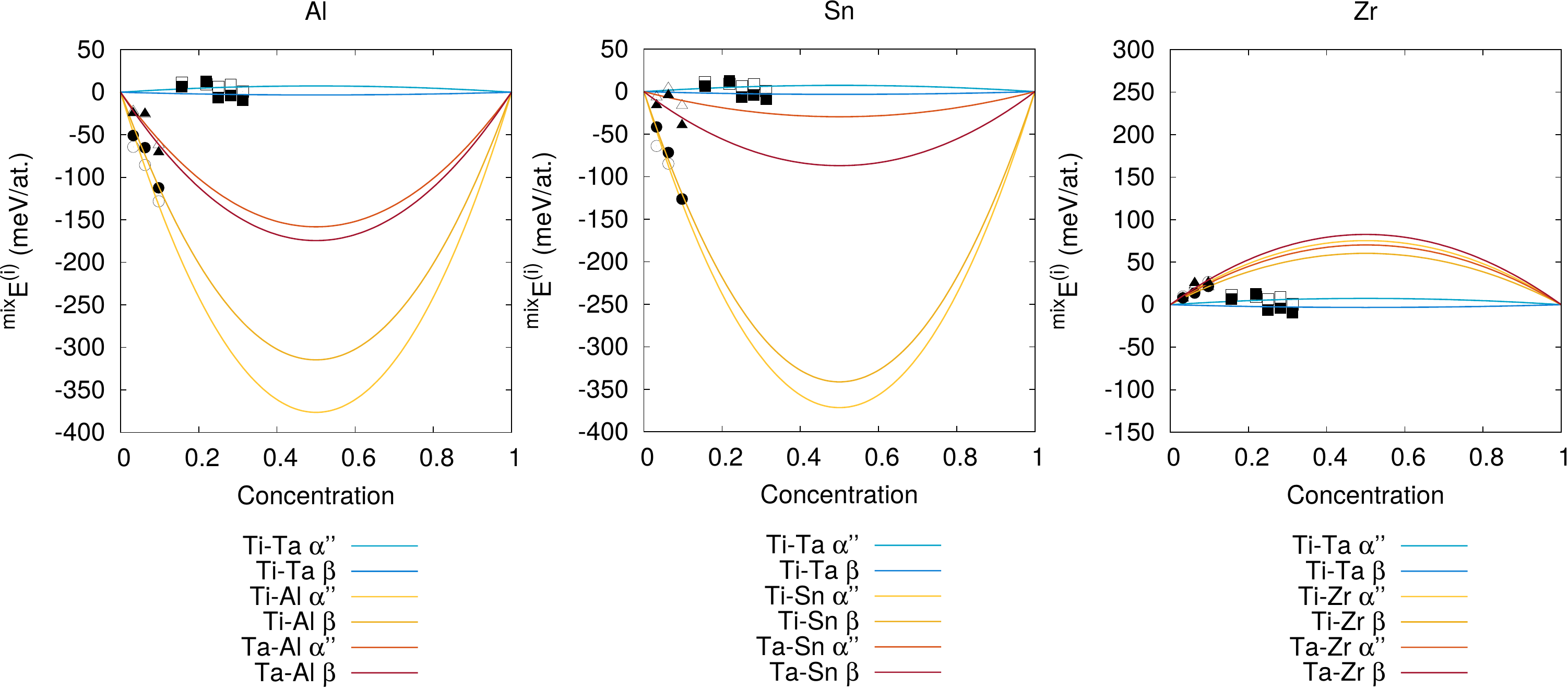}
\par\end{centering}
\caption{Mixing energies for Ti-Ta, Ti-Al, Ta-Al, Ti-Sn, Ta-Sn, Ti-Zr and Ta-Zr in the $\alpha''$ and $\beta$ phases as a function of the concentration of the second element in the compound. Error bars correspond to the size of the points.}
\label{binary_interac}
\end{figure*}

\setlength{\tabcolsep}{2pt}
\renewcommand{\arraystretch}{1.5}
\begin{table}
\begin{centering} 
\begin{tabular}{ccc}
\hline
 & $\alpha''$ & $\beta$\tabularnewline
 \hline
$k_\text{TiTa}$(meV/at.) & 30$\pm$10 & -10$\pm$20\tabularnewline
$k_\text{TiAl}$(meV/at.) & -1500$\pm$100 & -1300$\pm$100\tabularnewline
$k_\text{TaAl}$(meV/at.) & -630$\pm$80 & -700$\pm$120\tabularnewline
$k_\text{TiSn}$(meV/at.) & -1500$\pm$100 & -1400$\pm$100\tabularnewline
$k_\text{TaSn}$(meV/at.) & -120$\pm$80 & -350$\pm$120\tabularnewline
$k_\text{TiZr}$(meV/at.) & 300$\pm$5  & 240$\pm$5\tabularnewline
$k_\text{TaZr}$(meV/at.) & 280$\pm$50 & 330$\pm$50\tabularnewline
\hline
\end{tabular}
\par
\end{centering}
\caption{Fitted $k$ parameters obtained from the curvature of the mixing energies for the considered binary compounds with statistical errors.}
\label{curvatures} 
\end{table}

\setlength{\tabcolsep}{2pt}
\renewcommand{\arraystretch}{1.5}
\begin{table*}
\begin{centering} 
\begin{tabular}{ccccc}
\hline
 & Ti-Ta-Al (d.f.) & Ti-Ta-Al (b.i.) & Ti-Ta-Sn (b.i.) & Ti-Ta-Zr (b.i.)\tabularnewline
\hline
A (K/at.\%) & -19.6$\pm$0.8 & -23.9$\pm$0.6 & -23.9$\pm$0.6 & -23.9$\pm$0.6\tabularnewline
B (K/at.\%) & 27$\pm$16 & 24$\pm$20 & 6$\pm$20 & -9$\pm$1\tabularnewline
C (K/(at.\%)$^{2}$) & -1.7$\pm$0.6 & -0.4$\pm$0.5 & -0.4$\pm$0.4 & 0.1$\pm$0.1\tabularnewline
D (K) & 720$\pm$20 & 940$\pm$60 & 940$\pm$60 & 940$\pm$60\tabularnewline
\hline
\end{tabular}
\par
\end{centering}
\caption{Coefficients in eq.~\eqref{fitting_function} calculated using eq.~\eqref{system} with statistical errors. Comparison between direct fitting (d.f.) and binary interaction method (b.i.).}
\label{fit_binary} 
\end{table*}

To understand better the meaning of the coefficients A, B, C, and D it is possible to reformulate eq.~\eqref{fitting_function} in terms of binary interactions using the mixing energy.
Comparing eq.~\eqref{II_order} and \eqref{fitting_function}, the coefficients A, B, C, and D can be expressed as a function of $\Delta k_\text{nm}$ and $\lambda_\text{n}$ 
\begin{align}
\label{system}
\text{A}&=\lambda_\text{Ta}-\lambda_\text{Ti}\\ \nonumber
\text{B}&=\Delta k_\text{TiX}+\lambda_\text{X}-\lambda_\text{Ti}\\ \nonumber
\text{C}&=\Delta k_\text{TaX}-\Delta k_\text{TiX}\\ \nonumber
\text{D}&=\lambda_\text{Ti} 
\end{align}
The advantage of this formulation is that $\lambda_\text{n}$ can be calculated with DFT from the energy of the pure elements in the $\alpha''$ and $\beta$ phases and $k_\text{nm}$ can be fitted from the mixing energies of binary n-m systems.
This makes the interpretation of the coefficients more transparent and allows to assess how the different atomic interactions contribute to the formation energy of Ti-Ta-X.
The method is also computationally less demanding than directly fitting a set of DFT data using eq.~\eqref{fitting_function} because unary and binary systems are remarkably easier to be treated within DFT than ternary systems.  
In addition, the simultaneous fitting of 4 parameters requires many more data points at different concentrations for overdetermination of the parameters.
\newline
The total energy of element $n$ in the $i$-th phase, $E_\text{n}^{\text{(i)}}$, and the coefficients $k_\text{TiX}^\text{(i)}$ and $k_\text{TaX}^\text{(i)}$ were extracted from DFT calculations with the high accuracy setting.
32-atom binary SQS with  3.125\%, 6.25\% and 9.375\% X content, respectively, were generated and the lattice sites were populated with Ti and Al, Ta and Al, Ti and Sn, Ta and Sn, Ti and Zr and Ta and Zr, respectively.
The cell shape and volume as well as the atomic positions were fully relaxed for Ti-Al in both phases, Ta-Al in the $\beta$ phase, Ti-Sn in the $\alpha''$ phase, Ta-Sn and Ta-Zr in the $\beta$ phase. 
The other structures do not correspond to local minima of the energy and therefore relax to different configurations upon full relaxation.
In these cases the atomic positions of the X element, the volume and for $\alpha''$ the ratios of the lattice parameters were relaxed.
Random binary Ti-X and Ta-X in the $\alpha''$ and $\beta$ structures are indeed ``artificial'' compounds, which do not exist in nature, but can nevertheless provide insight into the chemical interactions occurring in Ti-Ta-X. 
\newline
The mixing energies  for these compounds as a function of composition are shown in fig.~\ref{binary_interac}. 
The data were fitted with second order polynomials according to eq.~\eqref{random_all} to determine $k_\text{TiX}^\text{(i)}$ and $k_\text{TaX}^\text{(i)}$.
A statistical error of 7.5 meV/at.~was assumed in evaluating eq.~\eqref{mixing}.
All fits were acceptable by means of the $\chi^2$ test.
The fitted $k$ parameters obtained from the curvatures are reported in tab.~\ref{curvatures}. 
For comparison, the mixing energy of Ti-Ta is also included in fig.~\ref{binary_interac} using the data for binary Ti-Ta from the previous Section.
As already pointed out in Section~\ref{model}, the mixing energy for Ti-Ta is negligible, since $k^{(\alpha'')}_\text{TiTa}$ and $k^{(\beta)}_\text{TiTa}$ are very small.
In contrast, the mixing energies of Ti-Al and Ti-Sn exhibit strong, positive curvatures in both phases resulting in large, negative values of $k$.  This means that the chemical reactions
\[
\text{Ti}^{(\alpha'',\,\beta)}+\text{Al}^{(\alpha'',\,\beta)}\rightarrow \text{TiAl}^{(\alpha'',\,\beta)}
\]
and
\[
\text{Ti}^{(\alpha'',\,\beta)}+\text{Sn}^{(\alpha'',\,\beta)}\rightarrow \text{TiSn}^{(\alpha'',\,\beta)}
\]
are exothermic, because Ti, Al and Sn are very unstable in both $\alpha''$ and $\beta$.
The curvatures for Ta-Al and Ta-Sn are smaller, but still positive, i.e.~also in this case a mixing is favourable.
Ti-Zr and Ta-Zr have instead relatively small, negative curvatures, since Zr is isoelectronic to Ti and little heat is gained or lost upon alloying.
\newline
In the binary interaction model, the curvatures from tab.~\ref{curvatures} are used to estimate the coefficients A, B, C, and D in eq.~\eqref{fitting_function} using eq.~\eqref{system}.
The corresponding coefficients are compiled in tab.~\ref{fit_binary}.
For Ti-Ta-Al, the coefficients extracted from the binary interaction model are overall in good agreement with those determined by direct fitting of the data from ternary alloy configurations.
The coefficient C is somewhat underestimated, which may be due to that fact that the driving forces towards ordering are stronger for binary alloys making the fitting of the curvatures slightly biased. 
Furthermore, the direct fitting of DFT or experimental data using eq.~\eqref{fitting_function} introduces  contributions from higher order terms in the coefficient C, while eq.~\eqref{II_order} and \eqref{system} hold strictly up to second order in the concentrations.
Nevertheless, the competition between the coefficients B and C, which is responsible for the change in the slope of the transformation temperatures and $\Delta E^{\beta-\alpha''}$ discussed in Sections~\ref{experimental} and~\ref{dir_fit}, can be interpreted in terms of binary interactions.
The coefficient B is determined by the interaction between Ti and Al and by the total energies of Ti and Al in the two phases (eq.~\eqref{system}).
Ti-Al is found to stabilize the $\alpha''$ over the $\beta$ phase (the curvature of the mixing energy of $\alpha''$ is larger than that of $\beta$), while pure Al and pure Ti are not found to exhibit a preference for any of the two phases.
This means that overall the $\alpha''$ phase will be lower in energy and thus $\text{B}>0$.
On the other hand, the C coefficient is determined by $\Delta k_\text{TaAl}-\Delta k_\text{TiAl}$ (eq.~\eqref{system}).
Since the difference in the curvatures of mixing energies for $\beta$ and $\alpha''$ is larger in Ti-Al  than  in Ta-Al, one has $\text{C}<0$.
\newline
Essentially, the inversion of the trend in the transformation temperatures in Ti-Ta-Al arises from
\begin{align}
\label{condition}
 \Delta k_\text{TiAl} &\gg -\lambda_\text{Al}+\lambda_\text{Ti}\\ \nonumber
 \Delta k_\text{TiAl}&>\Delta k_\text{TaAl}
\end{align}
that is, Ti-Al favours considerably the $\alpha''$ phase over the $\beta$ phase.
The criterion \eqref{condition} can be regarded as a simple (not unique) condition for Ti-Ta-X to exhibit the same inversion of the trend in the transformation temperatures as observed for Ti-Ta-Al and it may eventually be exploited to identify other alloying elements that produce the same change of trend. 
\newline
From tab.~\ref{fit_binary} it can be seen that the coefficient B in Ti-Ta-Sn is much smaller than in Ti-Ta-Al, which is due to the fact that $\Delta k_\text{TiSn} \simeq -\lambda_\text{Sn}+\lambda_\text{Ti}$ and therefore the first condition in eq.~\eqref{condition} does not hold.
This is presumably why in Ti-Ta-Sn the transformation temperatures decrease for increasing $c_\text{Sn}$ at every $c_\text{Ta}$, as found experimentally by Kim et al.~\cite{Kim2011}.
\newline
For Ti-Ta-Zr, the situation is somehow different, because of the very different valence electronic configuration of Zr with respect to the $p$-valent metals Al and Sn. 
From the analysis of tab.~\ref{curvatures} and fig.~\ref{binary_interac}, it can be deduced that $\Delta k_\text{TiZr} \simeq \Delta k_\text{TaZr} \simeq 0$ and thus $\text{C}\simeq0$.
Since $\text{B}<0$ and $\text{C}\simeq0$, no change in the trend for the transformation temperatures is predicted for Ti-Ta-Zr.
This agrees with the experimental data presented by Buenconsejo et al.~\cite{Buenconsejo2009a} and Zheng et al.~\cite{Zheng2013}.
\newline
In general, the predictions of the binary interaction model are in reasonable agreement with the experiments regarding the trend in the coefficients B and C when comparing Ti-Ta-Al, Ti-Ta-Sn and Ti-Ta-Zr.
However, it should be noted that the numerical values of the coefficients are affected by the approximations and biases already discussed earlier in this Section and the predictions by this model should be considered as qualitative indications. 
Nevertheless, the pairwise random approximation can be a great benefit to alloy design because of its simplicity and low computational effort.
It can, for example, be used in combination with a high-throughput search for stable SMAs to determine the alloy composition with highest martensite start temperature, as it allows to identify the criteria for the alloying elements X to maximize the transformation temperature.

\section{Conclusions}\label{conclusion}
In this work, an inversion of the trend in the transformation temperatures for increasing Al concentration has been measured experimentally in Ti-Ta-Al.
In contrast to the generally observed behaviour, according to which $M_\text{s}$ and $A_\text{s}$ decrease when Al is added, an increase has been detected.
A simple model based on the quasi-chemical approach has been developed to rationalize the compositional dependence of the transformation temperatures in ternary Ti-Ta-X alloys.
The parameters in this model were fitted directly for Ti-Ta-Al to DFT and experimental data which revealed the same inversion of the trend in $\Delta E^{\beta-\alpha''}$ and in the transformation temperatures.
The compositional dependence of $\Delta E^{\beta-\alpha''}$ in Ti-Ta-Al, which is directly related to the change in transformation temperatures, can be rationalized by considering only binary interactions.
From a random alloy approximation a criterion was extracted to predict the conditions for an inversion of the trend in the transformation temperatures.
The versatility induced by this inversion makes Ti-Ta-Al a promising candidate as HTSMAs.
This unique behavior is explained by the fact that Al stabilizes the $\alpha''$ phase when substituting Ta and that when $c_\text{Al}$ and $c_\text{Ta}$ are increased, the $\beta$ phase is stabilized.
\newline
The analytical model was subsequently extended to Ti-Ta-Sn and Ti-Ta-Zr, for which a qualitative correlation to the existing experimental data was provided.
In contrast to Ti-Ta-Al, these two alloys do not exhibit the inversion of the trend in the transformation temperatures.
\newline
The binary interactions method derived in this work provides a fast and elegant tool to obtain a qualitative understanding of the composition dependence of the transformation temperatures in ternary alloys and can be used as a guide for the development of high-temperature shape memory alloys.

\section*{Acknowledgements}
This work has been supported by Deutsche Forschungsgemeinschaft (DFG) within the research unit FOR 1766 (High Temperature Shape Memory Alloys, http://www.for1766.de), under the grant numbers FR2675/3-2 (sub-project 1) and RO3073/4-2 (sub-project 3).

\appendix
\section{Experimental data}\label{appendix_exp}

\begin{figure}
\begin{centering} 
\includegraphics[scale=0.7]{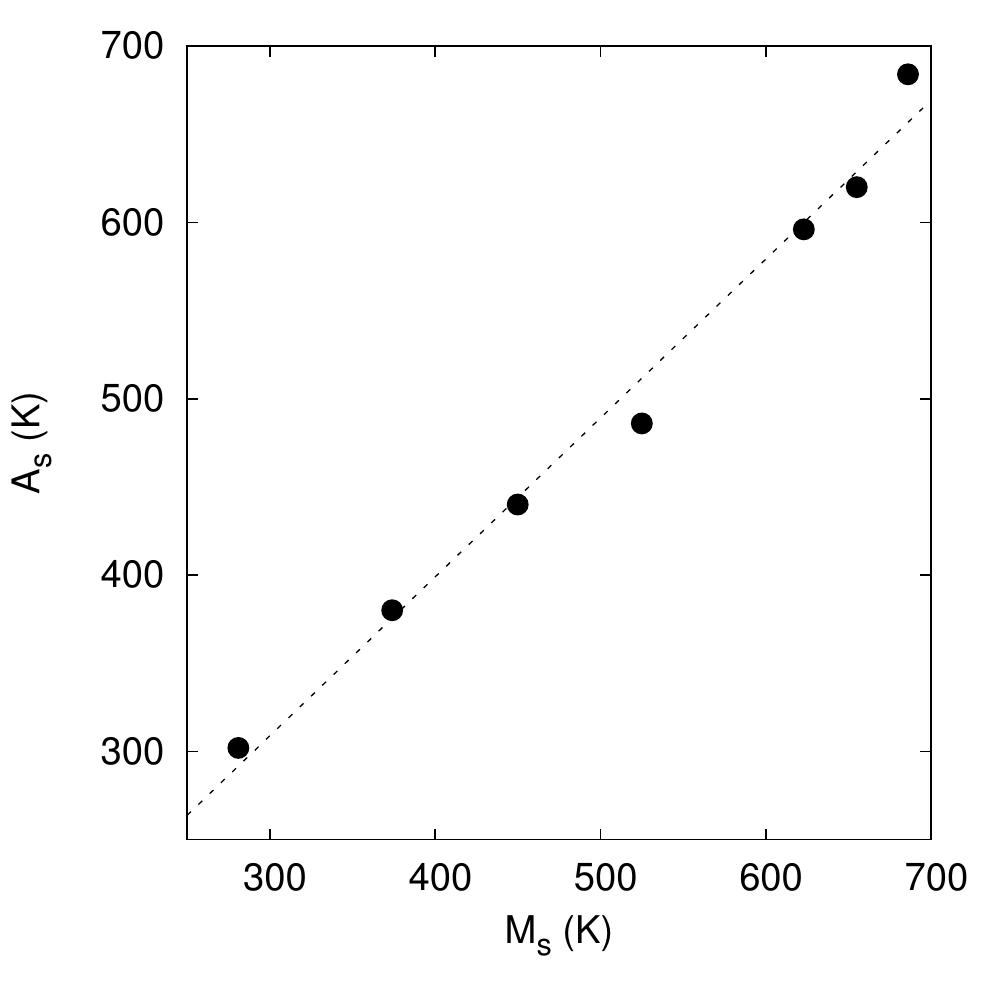}
\par\end{centering}
\caption{Austenite and martensite start temperatures for different alloy compositions.}
\label{Ms_As}
\end{figure}

Tab.~\ref{data_transf_temp} reports the experimental data on the transformation temperatures in Ti-Ta-Al.
Fig.~\ref{Ms_As} displays the corresponding austenite and martensite start temperatures measured in this work for different alloys.
The linear best fit shows that the temperatures are directly proportional to each other.

\begin{table}
\begin{centering} 
\begin{tabular}{ccccc}
\hline 
Alloy & $A_\text{s}$ (K) & $A_\text{f}$ (K) & $M_\text{s}$ (K) & $M_\text{f}$ (K)\tabularnewline
\hline 
Ti-16Ta & 760 & 792 & - & -\tabularnewline
Ti-16Ta-3Al & 764 & 797 & - & -\tabularnewline
Ti-16Ta-5Al & 803 & 831 & - & -\tabularnewline
Ti-16Ta-7Al & 835 & 868 & - & -\tabularnewline
Ti-20Ta & 684 & 695 & 686 & 660\tabularnewline
Ti-20Ta-3Al & 620 & 680 & 655 & 610\tabularnewline
Ti-20Ta-5Al & 596 & 615 & 623 & 577\tabularnewline
Ti-27.5Ta & 486 & 534 & 525 & 453\tabularnewline 
Ti-27Ta-3Al & - & - & 477 & -\tabularnewline
Ti-27Ta-5Al & - & - & 428 & -\tabularnewline
Ti-27Ta-7Al & - & - & 309 & -\tabularnewline
Ti-30Ta & 440 & 493 & 450 & 416\tabularnewline
Ti-30Ta-3Al & 380 & 408 & 374 & 313\tabularnewline
Ti-30Ta-5Al & 302 & 333 & 281 & -\tabularnewline
\hline
\end{tabular}
\par
\end{centering}
\caption{Transformation temperatures measured experimentally for Ti-Ta and Ti-Ta-Al.}
\label{data_transf_temp} 
\end{table}

\section{First principles data}\label{appendix_DFT}

Tab.~\ref{data_DFT} reports the first principles data on the energy difference between austenite and martensite in Ti-Ta-Al.

\begin{table}
\begin{centering} 
\begin{tabular}{cc|cc}
\hline 
Alloy & $\Delta E^{\beta-\alpha''}$(meV/at.) & Alloy & $\Delta E^{\beta-\alpha''}$(meV/at.)\tabularnewline
\hline 
Ti-28Ta-3Al & 8.4$\pm$1.0 & Ti-16Ta-3Al & 30.1$\pm$10.0\tabularnewline
Ti-25Ta-6Al & 12.7$\pm$1.0 & Ti-25Ta-3Al & 7.9$\pm$10.0\tabularnewline
Ti-13Ta & 28.4$\pm$5.0 & Ti-31Ta-3Al & 2.7$\pm$10.0\tabularnewline
Ti-16Ta & 44.1$\pm$5.0 & Ti-19Ta-6Al & 16.3$\pm$10.0\tabularnewline
Ti-22Ta & 39.2$\pm$5.0 & Ti-19Ta-6Al & 26.7$\pm$10.0\tabularnewline
Ti-25Ta & 16.8$\pm$5.0 & Ti-22Ta-6Al & 15.8$\pm$10.0\tabularnewline
Ti-28Ta & 10.3$\pm$5.0 & Ti-28Ta-6Al & -1.7$\pm$10.0\tabularnewline
Ti-31Ta & 7.2$\pm$5.0 & Ti-31Ta-6Al & -1.7$\pm$10.0\tabularnewline
\hline 
\end{tabular}
\par
\end{centering}
\caption{Formation energy differences between the $\beta$ and $\alpha''$ phases calculated with DFT for Ti-Ta and Ti-Ta-Al.
\label{data_DFT} 
}
\end{table}

\newpage

\bibliography{./biblio}

\end{document}